\documentclass{article}
\usepackage{epsfig}
\begin{document}

\begin{center}
{\LARGE \bf Heisenberg quantization for the }
\end{center}

\begin{center}
{\LARGE \bf systems  of identical particles and }
\end{center}

\begin{center}
{\LARGE \bf  the Pauli exclusion principle in  }
\end{center}
\begin{center}
{\LARGE \bf  noncommutative  spaces }
\end{center}
\vspace{1.00cm}
\begin{center}
{\LARGE\bf S. A. Alavi}
\end{center}

\begin{center}
\textit{High Energy Physics Division, Department of Physics,
University of
Helsinki\\
and\\
Helsinki Institute of Physics, FIN-00014 Helsinki, Finland.\\
On leave of absence from : Department of Physics, Ferdowsi
University of
Mashhad, Mashhad, P. O. Box 1436, Iran\\
E-mail: $Ali.Alavi@Helsinki.fi$\\
$s_{-}alialavi@hotmail.com$}
\end{center}

\textbf{1 Abstract}. \textsl{We study the Heisenberg quantization
for the systems of identical particles in noncommtative spaces. We
get fermions and bosons as a special cases of our argument, in the
same way as commutative case and therefore we conclude that the
Pauli exclusion principle is also valid in noncommutative spaces.}\\

\textbf{1 Introduction}.\\

Recently there have been notable studies on the formulation and
possible experimental consequences of extensions of the standard
(usual) quantum mechanics in the noncommutative spaces [1-13].
Many physical problems have been studied in the framework of
the noncommutative quantum mechanics (NCQM), see e.g. [1-4].\\
NCQM is formulated in the same way as the standard quantum
mechanics SQM (quantum mechanics in commutative spaces), that is
in terms of the same dynamical variables represented by operators
in a Hilbert space and a state vector that evolves according to
the Schroedinger equation :
\begin{equation}
i\frac{d}{dt}|\psi>=H_{nc}|\psi> .
\end{equation}
we have taken in to account $\hbar=1$ and $H_{nc}\equiv
H_{\theta}$ denotes the Hamiltonian for a given system in the
noncommutative space. In the literatures two
approaches have been considered  for constructing the NCQM :\\
a) $H_{\theta}=H$, so that the only difference between SQM and
NCQM is the presence of a nonzero $\theta$ in the commutator of
the position operators :
\begin{equation}
[\hat{x}_{i},\hat{x}_{j}]=i\theta_{ij}\hspace{1.cm}[\hat{x}_{i},\hat{p}_{i}]=i\delta
_{ij}\hspace{1.cm} [\hat{p}_{i},\hat{p}_{j}]=0 .
\end{equation}
b) By deriving the Hamiltonian from the Moyal analogue of the
standard Schroedinger equation :
\begin{equation}
i\frac{\partial}{\partial t}\psi(x,t)=H(p=\frac{1}{i}\nabla,x)\ast
\psi (x,t)\equiv H_{\theta}\psi(x,t) ,
\end{equation}
where $H(p,x)$ is the same Hamiltonian as in the standard theory,
and as we observe the $\theta$ - dependence enters now through the
star product [11]. In [13], it has been shown that these two
approaches lead to the same physical theory. For the Hamiltonian
of the type :
\begin{equation}
H(\hat{p},\hat{x})=\frac{\hat{p}^{2}}{2m}+V(\hat{x}) .
\end{equation}
the modified Hamiltonian $H_{\theta}$ can be obtained by a shift
in the argument of the potential [1,11] :
\begin{equation}
x_{i}=\hat{x}_{i}+\frac{1}{2}\theta_{ij}\hat{p}_{j}\hspace{2.cm}\hat{p}_{i}=p_{i} 
,
\end{equation}
which lead to
\begin{equation}
H_{\theta}=\frac{p^{2}}{2m}+V(x_{i}-\frac{1}{2}\theta_{ij}p_{j}) .
\end{equation}
The variables $x_{i}$ and $p_{i}$ now, satisfy in the same
commutation relations as the usual case :
\begin{equation}
[x_{i},x_{j}]=[p_{i},p_{j}]=0\hspace{1.cm}[x_{i},p_{j}]=\delta_{ij} .
\end{equation}
In this paper we study one of the most important problems in the
quantum mechanics namely the Heisenberg quantization for the
systems of identical particles and the Pauli exclusion principle
in noncommutative spaces. It appears that
the main difference is between one and two dimensions not between
two and three dimensions and therefore we study only one and two
dimensional spaces. Heisenberg quantization for the systems of
identical particles in the framework of SQM has been studied in
detailed in [14].The process of quantization in SQM consists of:
\\
a) Identifying the classical observables and the Lie algebra which
they form under poisson bracket.\\
b) Looking for the linear representations of this lie algebra such
that classical observables $A$ and $B$ are represented by
Hermitian operators $A$ and $B$ on some Hilbert space. \\
c)Replacing the Poisson bracket $C=\{A,B\}$ by the commutator :
\begin{equation}
\hat{C}=\frac{1}{i}[\hat{A},\hat{B}]=\frac{1}{i}(\hat{A}\hat{B}-\hat{B}\hat{A}) 
.
\end{equation}
The classical observables of a system of $N$ identical particles
are the real valued functions on the $N$ particles phase space.
Every observable must be symmetric as a function of the $N$
one-particle phase space variables, because if there would exist a
nonsymmetric observable, it can be used to distinguish beween
particles [15, 16]. Boson and fermion quantization are separate
possibilities, because the symmetry of the quantum mechanical
observables ensures that they do not mix symmetric and
antisymmetric wave functions. In order to apply the Heisenberg
quantization to a system of identical particles in NCQM, we follow
the same way as SQM  but we replace the commutator relations  (7)
by their noncommutative counterparts i.e. equ.(2). \\

\textbf{2 Two particles in one dimensional
spaces}\\

In this section We apply the Heisenberg quantization to systems of two
identical particles in commutative (usual) space. The fundemental
relations are the cononical commutation relations :
\begin{equation}
[x,x]=[p,p]=0 \hspace{2 cm} [x,p]=i .
\end{equation}
To describe the two-particles systems, we introduce the relative
coordinate and momentum [14]:
\begin{equation}
x=x_{(1)}-x_{(2)}\hspace{2 cm} p=\frac{1}{2}(p_{(1)}-p_{(2)}),
\end{equation}
where the subscript $\ell=1,2$ refers to the particles. If the two
particles are not identical, the relative $x$ and $p$ of the
two-particles system are observables and satisfy the same
cononical relations as (9), but if particles are identical, then
$x$ and $p$ are not observables any more, because they are
antisymmetric under exchange of the particles indices. The
observables can be [14]:
\begin{equation}
A=\frac{1}{4}(p^{2}+x^{2})\hspace{1 cm}
B=\frac{1}{4}(p^{2}-x^{2}) \hspace{1 cm} C=\frac{1}{4}(px+xp),
\end{equation}
which are symmetric under the exchange of the particles indices, i.e. 
$x\rightarrow -x$ and $p\rightarrow -p$ .
The observables $A$, $B$ and $C$ satisfy the $sp(1,R)$ algebra :
\begin{equation}
[A,B]=iC \hspace{1 cm}[A,C]=-iB \hspace{1 cm}[B,C]=-iA .
\end{equation}
We can transform the above commutation relations in to the more
familiar form by introducing the operators $B_{\pm}$ as follows :
\begin{equation}
B_{\pm}=B\pm iC .
\end{equation}
Then we have :
\begin{equation}
[A,B_{\pm}]=\pm B_{\pm} \hspace{2 cm} [B_{+},B_{-}]=-2A .
\end{equation}
We can define the irreducible
representation of the algebra by introducing an orthonormal vector
$|\alpha_{0},n>$, $n=0,1,2,...$ as follows [14] :
\begin{equation}
\Gamma |\alpha_{0},n>=\alpha_{0}(\alpha_{0}-1)|\alpha_{0},n>,
\end{equation}
\begin{equation}
A |\alpha_{0},n>=(\alpha_{0}+n)|\alpha_{0},n>,
\end{equation}
\begin{equation}
B_{+} |\alpha_{0},n>=\sqrt{(n+1)(n+2\alpha_{0})}|\alpha_{0},n+1>,
\end{equation}
\begin{equation}
B_{-} |\alpha_{0},n>=\sqrt{n(n-1+2\alpha_{0})}|\alpha_{0},n-1>,
\end{equation}
where $\Gamma=A^{2}-B^{2}-C^{2}$ is the Casimir operator and
$\alpha_{0}$ is the minimum eigenvalue of $A$(an arbitrary
constant which defines the representation). The cases
$\alpha_{0}=\frac{1}{4}$ and $\alpha_{0}=\frac{3}{4}$ correspond
to bosons and fermions respectively.

\textbf{3 Two identical particles in two dimensional non-commtative 
spaces.}\\

We first consider the problem in commutative spaces [14]. We can
describe the two particles systems by relative coordinates in the
same way as in one dimensional case. We define complex quantities
$a_{j\pm}$ as follows :
\begin{equation}
a_{j\pm}=\frac{1}{\sqrt{2}}(p_{j}\pm i x_{j}),\hspace{2 cm}j=1,2.
\end{equation}
They satisfy in the following commutation relations :
\begin{equation}
[a_{j+},a_{k+}]=[a_{j-},a_{k-}]=0 \hspace{1
cm}[a_{j-},a_{k+}]=\delta_{jk} .
\end{equation}
As we mentioned before, $x$ and $p$ are not observables. The
generalized one-dimensional observables $A$, $B$ and $C$ are :
\begin{equation}
A_{j}=\frac{1}{4}(a_{j+}a_{j-}+a_{j-},a_{j+}),\hspace{2 cm}
B_{j\pm}=B_{j}\pm i C_{j}=\frac{1}{2}(a_{j\pm})^{2}.
\end{equation}
In addition we have two-dimensional observables which are the real
and imaginary parts of :
\begin{equation}
D_{\pm}=D_{re}\pm i D_{im}=a_{1\pm} a_{2\pm} \hspace{2 cm}
E_{\pm}=E_{re}\pm i E_{im}=a_{1\mp}\mp a_{2\pm} .
\end{equation}
There are two $sp(1,R)$ algebras $A_{1}$, $B_{1\pm}$ and $A_{2}$,
$B_{2\pm}$ [14] :
\begin{equation}
[A_{j},B_{j\pm}]=\pm B_{j\pm} \hspace{1.5
cm}[B_{j-},B_{j+}]=2A_{j}\hspace{.5 cm} j=1,2 .
\end{equation}

There are also two other algebras, one sp(1,R) algebra :
\begin{equation}
[A_{1}+A_{2},D_{\pm}]=\pm D_{\pm} \hspace{2
cm}[D_{+},D_{-}]=-2(A_{1}+A_{2}) .
\end{equation}
and one $su(2)$ algebra :
\begin{equation}
[A_{2}-A_{1},E_{\pm}]=\pm E_{\pm} \hspace{2
cm}[E_{+},E_{-}]=2(A_{2}-A_{1}) .
\end{equation}

Now we study the same problem in two dimensional noncommutative
spaces. The operators $a_{j\pm}$ are given by :
\begin{equation}
\hat{a}_{j\pm}=\frac{1}{\sqrt{2}}(\hat{p}_{j}\pm i \hat{x}_{j}) ,
\end{equation}
where $\hat{p}_{j\pm}$ and $\hat{x}_{j}$ satisfy in the Heisenberg
commutation relations in noncommutative spaces i.e. equ.(2). We
have the following commutation relations :
\begin{equation}
[\hat{a}_{j+},\hat{a}_{k+}]=[\hat{a}_{j-},\hat{a}_{k-}]=-\frac{i}{2}\theta_{jk},\hspace{1
cm}[\hat{a}_{j-},\hat{a}_{k+}]=\delta_{ij}+\frac{i}{2}\theta_{jk} .
\end{equation}

The two $sp(1,R)$ algebras (23) are also valid in this case, but
the commutation relations (24) and (25) are no longer valid and in
addition $A_{1}$ and $A_{2}$ don't commute with each other and can
not have common eigenvectors. Now we use the variables $x_{i}$ and
$p_{i}$ introduced in equ.(5), then we have :
\begin{equation}
a_{j\pm}=\frac{1}{\sqrt{2}}(p_{j}\pm i x_{j}),\hspace{2 cm}j=1,2 .
\end{equation}
where $x_{\ell}$ and $p_{\ell}$ satisfy in (7). One can show that
$a_{j\pm}$ satisfy the same commutation relations as commutative
case :
\begin{equation}
[a_{j+},a_{k+}]=[a_{j-},a_{k-}]=0 \hspace{1
cm}[a_{j-},a_{k+}]=\delta_{ij},
\end{equation}
and therefore the operators $A_{j}$, $B_{j\pm}$, $A_{2}\pm A_{1}$,
$E_{\pm}$ and $D_{\pm}$ satisfy in the same $sp(1,R)$ and $su(2)$
algebras as commutative space :
\begin{equation}
[A_{j},B_{j\pm}]=\pm B_{j\pm} \hspace{1.5
cm}[B_{j-},B_{j+}]=2A_{j}\hspace{.5 cm} j=1,2 ,
\end{equation}
\begin{equation}
[A_{1}+A_{2},D_{\pm}]=\pm D_{\pm} \hspace{2
cm}[D_{+},D_{-}]=-2(A_{1}+A_{2}) ,
\end{equation}
\begin{equation}
[A_{2}-A_{1},E_{\pm}]=\pm
E_{\pm}\hspace{2.cm}[E_{-},E_{+}]=-2(A_{2}-A_{1}) .
\end{equation}
We also have the following commutation relations :
\begin{equation}
[A_{1},A_{2}]=0 , \\
\end{equation}
\begin{equation}
[E_{-},D_{+}]=2B_{1+} \hspace{2.cm} [D_{-},E_{+}]=2B_{1-} ,
\end{equation}
\begin{equation}
[E_{+},D_{+}]=2B_{2+}\hspace*{2cm}[D_{-},E_{-}]=2B_{2-} ,
\end{equation}
\begin{equation}
[E_{-},B_{2+}]=[E_{+},B_{1+}]=D_{+}   ,
\end{equation}
\begin{equation}
[E_{-},B_{1-}]=[E_{+},B_{2-}]=-D_{-}  ,
\end{equation}
\begin{equation}
[D_{-},B_{1+}]=-[D_{+},B_{2-}]=E_{-}  ,
\end{equation}
\begin{equation}
[D_{-},B_{2+}]=-[D_{+},B_{1-}]=E _{+} .
\end{equation}

We observe that by using the variables $x_{i}$ and $p_{i}$ defined by
equ.(5), all of the commutation relations are the same as
commutative case
and the algebra governs physics of the system of identical particles and its 
representation remain unchanged.\\
The operators $A_{1}$ and $A_{2}$ together with the rasing and
lowering operators $B_{1\pm}$, $B_{2\pm}$, $D_{\pm}$ and $E_{\pm}$
form a cartan basis for $sp(2,R)$ algebra.\\
The "symplectic" algebra $sp(2,R)$ has 10 generators $a_{i}$,
which are $4\times 4$ matrices and satisfy in the following relation :
\begin{equation}
\tilde{a}_{i} J=-J a_{i},
\end{equation}
where $\tilde{a}_{i}$ is the transpose of the $a_{i}$, and the
matrix $J$ is defined by : \\

\hspace{4.5cm}$ J = \left[\begin{array}{cc}
0 & {\bf 1} \\
-{\bf 1} & 0
\end{array} \right ]$\\

where $\textbf{1}$ is the $2\times 2$ unit matrix. The graphic
representation of the root vectors of $sp(2,R)$ algebra can be
drawn by introducing the angle $\phi$ between the root vectors
$\alpha$ and $\beta$ :
\begin{equation}
\cos(\phi)=\frac{(\alpha,\beta)}{\sqrt{(\alpha,\alpha)(\beta,\beta)}} ,
\end{equation}
where $(\alpha,\beta)=\alpha^{i}\beta_{i}$ is the scalar product.
Fig.1 shows the graphic representation of the root
vectors.\\
\begin{figure}
\centering \epsfig{file=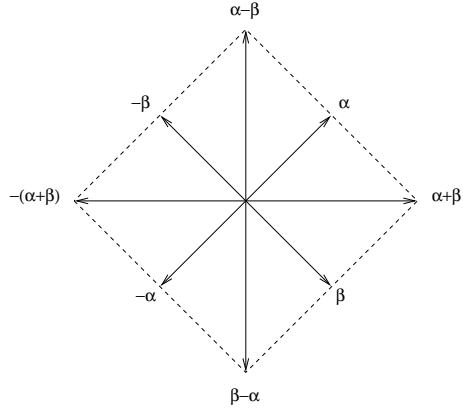,angle=0,width=0.5\textwidth}
\caption{Vector diagram of the lie algebra $C_2=sp(2,{\bf R})$ (according to
Cartan's notation). It contains 10 vectors including the two null
vectors in the center. }
\label{Fig1}
\end{figure}
Like commutative case we can construct an arbitrary representation
of the $sp(2,{\bf R})$ by a common eigenvector of $A_{1}$ and $A_{2}$ :
\begin{equation}
A_{1}|a>=a_{1}|a> \hspace{2 cm} A_{2}|a>=a_{2}|a> ,
\end{equation}
where $(a_{1},a_{2})$ is a weight of the representation. All of
the operators $B_{1-}, B_{2-}, D_{-}$ and $E_{\pm}$ lower the
eigenvalues of at least one of the $A_{1}$ or $A_{2}$ by either
$\frac{1}{2}$ or $1$, and we can use this fact to obtain the
eigenvectors and eigenvalues of $A_{1}$ and $A_{2}$, we will have
[14] :
\begin{equation}
A_{1}|jkmn>=(a_{10}+j+\frac{m-n}{2})|jkmn> ,
\end{equation}
\begin{equation}
A_{2}|jkmn>=(a_{20}+k+\frac{m+n}{2})|jkmn> ,
\end{equation}
where $(a_{10},a_{20})$ are the weights of the lowest leading weight vector
$|0>$.\\
Like in one dimension we can get bosons and fermions as special
cases of the Heisenberg quantization in two dimensions : \\
a) If $a_{10}=a_{20}=\frac{1}{4}$, the eigenfunction is symmetric
in both $x_{1}$ and $x_{2}$ or antisymmetric in both and this
corresponds to the bosons.\\
b) If $a_{10}=\frac{3}{4}$, $a_{20}=\frac{1}{4}$, the
eigenfunction has opposite symmetry in
$x_{1}$ and $x_{2}$ which corresponds to fermions, thus :\\

\hspace{3.cm}$2(a_{10} - a_{20})$ = K = $\left\{ \begin{array}{ll}
0 & \mbox{for Bosons} \\
1 & \mbox{for Fermions}
\end{array} \right. \ $\\

This means that a system of two identical fermions or bosons in
two dimensional noncommutative spaces has a definite symmetry in
each of the relative coordinates $x_{1}$ and $x_{2}$ (like
commutative case) and therefore the Pauli exclusion principle is
also valid in noncommutative spaces.\\

\textbf{Acknowledgment.}\\
I would like to thank Professor P. Pre\v{s}najder for his valuable comments. 
I am also very grateful to Professor Masud Chaichian for his warm
hospitality during my visit to the University of Helsinki. I
acknowledge Dr. S. M. Harun-or-Rashid for his help on preparation
of the manuscript. This work is partialy supported by the Ministry
of Scince, Research and Technology of Iran.\\

\textbf{References.}\\

1. M. Chaichian, M. M. Sheikh-Jabbari and A. Tureanu, Phys. Rev.
Lett. 86, 2716 (2001).\\
2. M. Chaichian, A. Demichev, P. Pre$\breve{s}$najder, M. M.
Sheikh-Jabbari, and A. Tureanu, Phys. Lett. B 527 (2002) 149.\\
3. A. Smailagic and E. Spallucci, Phys. Rev. D 65, 107701
(2002).\\
4. $\ddot{O}$. Dayi and A. Jellal, Phys. Lett. A 287, 349
(2001).\\
5. P-M Ho and H-C Kao, Phys. Rev. Lett 88, 151 602 (2001).\\
6. R. Banerjee, hep-th/0106280. \\
7. J. Gamboa, F, M$\acute{e}$ndez, M. Loewe and J. C. Rojac, Mod.
Phys. Lett. A 16, 2075 (2001).\\
8. V. P. Nair and A. P. Polychronakos, Phys. Lett. B 505, 267
(2001).\\
9. B. Muthukamar and R. Mitra, hep-th/0204149.\\
10. D. Kochan and M. Demetrian, hep-th/0102050.\\
11. L. Mezincesu, hep-th/0007046.\\
12. Y. Zunger, JHEP o1o4, 039 (2001).\\
13. O. Espinosa and P. Gaete, hep-th/0206066.\\
14. J. M. Leinaas and J. Myrheim, International Journal of Modern
Physics A, Vol.8, No.21 (1993) 3649.\\
15. W. Heisenberg, Z. Phys. 38, 411 (1926).\\
16. P. A. M. Dirac, Proc. R. Soc. London A 112, 661 (1926).\\
\end{document}